\documentstyle[eqsecnum,aps]{revtex}

\newcommand{\hn}{\hat \nabla}

\newcommand{\gab}{g_{ab}}

\newcommand{\hTab}{\hat T_{ab}}

\newcommand{\hGab}{\hat G_{ab}}

\begin{document}

\draft

\title{Brans-Dicke-type theories and avoidance of the cosmological singularity}

\author{ Israel Quiros\thanks{israel@uclv.etecsa.cu}, Rolando Bonal and Rolando Cardenas}
\address{ Departamento de Fisica. Universidad Central de Las Villas. Santa Clara. CP: 54830 Villa Clara. Cuba }

\date{\today}

\maketitle

\begin{abstract}

We study flat Friedmann-Robertson-Walker cosmology in Brans-Dicke-type theory of gravitation with minimal coupling between the scalar field and the matter fields in the Einstein frame (general relativity with an extra scalar field) for arbitrary values of the Brans-Dicke parameter $\omega>-\frac{3}{2}$. It is shown that the cosmological singularity occuring in the Einstein frame formulation of this theory is removed in the Jordan frame in the range $-\frac{3}{2}<\omega\leq -\frac{4}{3}$. This result is interpreted in the ligth of a viewpoint (first presented in reference \cite{iq}) asserting that both Jordan frame and Einstein frame formulations of general relativity are physically equivalent. The implications of the obtained result for string theory are outlined.

\end{abstract}


\section{Introduction}

Scalar-tensor(ST) theories of gravity can be formulated in an infinite number of equivalent frames related by conformal rescalings of the spacetime metric. Among all conformally related frames the Jordan frame(JF) and the Einstein frame (EF) are distinguished \cite{sok}.

Although the JF and the EF formulations of a given ST theory provide mathematically equivalent descriptions of the same physics, the physical equivalence of these descriptions is under discussion (for an exhaustive review on the subject see \cite{fgn}). Moreover, most authors on the subject share the conviction that only one of the conformally related frames is the 'physical frame'. Other admit that JF and EF formulations of ST theory provide just two different descriptions of the same physics but, they claim, only one of both JF and EF metrics is the 'physical metric', i.e. the metric that is measured with clocks and rods made of ordinary matter \cite{ss}.

Among those that share the viewpoint of the non-physical equivalence of the JF and the EF formulations, it does not exists a unified criterion about which frame is the physical one \cite{sok,fgn}. Some authors of this group choose the Jordan frame as the physical frame since the ordinary matter is minimally coupled to the JF metric. Other reject this choice using energy arguments. The scalar field kinetic energy is negative definite, or indefinite in the Jordan frame. This implies that, in this frame, the theory does not have a stable ground state\cite{fgn}. In the Einstein frame, meanwhile, the scalar field possesses a positive definite kinetic energy for $\omega>-\frac{3}{2}$. We feel this remains an open question.

In reference \cite{iq} a point of view on this subject was presented that avoides any discussion on the physical preference of one or another conformal frame for the formulation of a given theory of gravity. It is based on the following observation. The conformal transformations of the metric can be considered as particular transformations of the units of length, time, and mass (a simple point-dependent scale factor applied to the units of length, time, and reciprocal mass)\cite{dk}. Spacetime coincidences are not affected by these transformations, i.e. spacetime measurements are not sensitive to the conformal rescalings of the metric\footnotemark\footnotetext{Another way of looking at this is realizing that the experimental measurements deal always with dimensionless numbers and these are unchanged under the transformations of the physical units. For a readable discussion on the dimensionless nature of measurements we recommend reading section II of reference \cite{am}}. This means that experimental observations are unchanged under these transformations. Consequently, the different conformal formulations of a given theory of gravity are experimentally indistinguishable. This line of reasoning leads that a statement such like 'the Jordan frame (or any other) formulation of Brans-Dicke-type theories is the physical one' is devoid of any physical, i.e. experimentally testable meaning. It can be taken as an independent postulate of the theory only. Then the discussion about which conformal frame is the physical one is devoid of interest since it is a non-well-posed question.

An alternative approach to this subject can be based on the following postulate. The different conformal representations of a given BD-type theory of gravity are physically equivalent\cite{iq}.

In the present paper we shall study the consequences of this postulate for flat Friedmann-Robertson-Walker(FRW) cosmology in Brans-Dicke-type theory of gravitation with minimal coupling between the scalar field and the matter fields in the Einstein frame. This last is just general relativity(GR) with an extra scalar field.

This paper has been organized as follows. In section II the class of Brans-Dicke-type theories of gravitation is presented and a brief discussion on conformal equivalence among members of this class is given. The meaning of the physical equivalence among the different conformal formulations is discussed. The Jordan frame formulation of general relativity is considered in section III. The Jordan frame flat FRW cosmology for a universe filled with barotropic perfect-fuid-type matter is studied in section IV. The implications for string theory of the results obtained in this last section are outlined in section V. Finally, in section VI a physical discussion of these results is given and the final fate of the cosmological singularity is conjectured.

\section{Brans-Dicke-type theories of gravity and conformal equivalence}

The Jordan frame Lagrangian for BD-type theories is given by:

\begin{equation}
L=\frac{\sqrt{-g}}{16\pi}(\phi R -\frac{\omega}{\phi}(\nabla\phi)^2),
\end{equation}
where $R$ is the Ricci scalar of the Jordan frame metric $\bf g$, $\phi$ is the BD scalar field and $\omega$ is the BD coupling constant (a free parameter). 

Under the conformal rescaling of the metric:

\begin{equation}
\hat g_{ab}=\phi g_{ab},
\end{equation}
and the scalar field redefinition $\hat \phi=ln \phi$, the JF Lagrangian for BD-type theory (2.1) is mapped into the Einstein frame Lagrangian for BD-type theory:

\begin{equation}
L=\frac{\sqrt{-\hat g}}{16\pi}(\hat R -(\omega+\frac{3}{2})(\hat \nabla \hat \phi)^2),
\end{equation}
where $\hat R$ is the curvature scalar given in terms of the EF metric $\hat {\bf g}$. 

Respecting interactions with matter in BD-type theory, only two possibilities seem to be physically interesting and reasonable \cite{sok}:

1. Matter minimally couples to the metric in Jordan frame:

\begin{equation}
L=\frac{\sqrt{-g}}{16\pi}(\phi R -\frac{\omega}{\phi}(\nabla\phi)^2)+L_{matter},
\end{equation}
where $L_{matter}$ is the Lagrangian density for the matter fields. Theory given by (2.4) is just the JF formulation of Brans-Dicke theory\cite{bdk}. 

2. Matter minimally couples to the metric in the Einstein frame:

\begin{equation}
L=\frac{\sqrt{-\hat g}}{16\pi}(\hat R -(\omega+\frac{3}{2})(\hat \nabla \hat \phi)^2)+L_{matter}.
\end{equation}

In this case the scalar field $\hat \phi$ is minimally coupled to curvature so the dimensional gravitational constant $G$ is a real constant. Due to the minimal coupling between ordinary matter and the spacetime metric the rest mass of any test particle $m$ is constant too over the manifold. This leads that the dimensionless gravitational coupling constant $Gm^2$ ($\hbar=c=1$) is a real constant too unlike BD theory where $Gm^2\sim \phi^{-1}$. 

The field equations derivable from (2.5) are:

\begin{equation}
\hGab=8\pi\hTab+(\omega+\frac{3}{2})(\hn_a \hat \phi \hn_b \hat \phi- \frac{1}{2} \hat g_{ab} (\hn \hat \phi)^2),
\end{equation}

\begin{equation}
{\hat {\Box}} \hat \phi=0,
\end{equation}
with $\hat G_{ab} \equiv \hat R_{ab}-\frac{1}{2} \hat g_{ab} \hat R$.

The conservation equation:

\begin{equation}
\hat \nabla_n \hat T^{na}=0,
\end{equation}
is fulfilled. $\hat T_{ab}=\frac{2}{\sqrt{-\hat g}}\frac{\partial}{\partial \hat g^{ab}}(\sqrt{-\hat g} L_{matter})$ are the components of the stress-energy tensor for the ordinary matter fields. 

The theory given by (2.5) is just Einstein's general relativity with a scalar field as an additional matter source of gravity. For $\hat \phi=const.$ or $\omega=-\frac{3}{2}$ we recover the usual Einstein's theory. This formulation of general relativity is linked with Riemann geometry because the test particles follow the geodesics of the metric $\bf \hat g$,

\begin{equation}
\frac{d^2x^a}{d\hat s^2}=-\hat \Gamma ^a_{mn} \frac{dx^m}{d\hat s} \frac{dx^n}{d\hat s},
\end{equation}
where $\hat \Gamma ^a_{bc}=\frac{1}{2} \hat g^{an}(\hat g_{bn,c}+\hat g_{cn,b}-\hat g_{bc,n})$ are the Christoffel symbols of the metric $\bf \hat g$. In fact, Riemann geometry is based on the parallel transport law:

\begin{equation}
d\xi^a=-\hat \gamma^a_{mn}\xi^m dx^n,
\end{equation}
and the length preservation requirement:

\begin{equation}
d\hat g(\xi,\xi)=0,
\end{equation}
where, in the coordinate basis $\hat g(\xi,\xi)=\hat g_{nm}\xi^n\xi^m$, $\hat \gamma^a_{bc}$ are the affine connections of the manifold, and $\xi^a$ are the components of an arbitrary vector $\bf {\xi}$. The above postulates of parallel transport and length preservation lead that, in Riemann geometry, the affine connections of the manifold coincide with the Christoffel symbols of the metric $\hat \gamma^a_{bc}=\hat \Gamma^a_{bc}$. Under the conformal transformation (2.2) the Lagrangian (2.4) is mapped into the Einstein frame Lagrangian for BD theory:

\begin{equation}
L=\frac{\sqrt{-\hat g}}{16\pi}(\hat R -(\omega+\frac{3}{2})(\hat \nabla \hat \phi)^2)+e^{-2\hat \phi}L_{matter},
\end{equation}
while (2.5) is mapped into:

\begin{equation}
L=\frac{\sqrt{-g}}{16\pi}(\phi R + \phi^{-1}(\nabla\phi)^2)+\phi^2L_{matter},
\end{equation}
that is the JF Lagrangian for general relativity with an extra scalar field. At the same time, under (2.2) the parallel transport law (2.10) is mapped into:

\begin{equation}
d\xi^a=-\gamma^a_{mn}\xi^m dx^n,
\end{equation}   
where $\gamma^a_{bc}=\Gamma^a_{bc}+\frac{1}{2} \phi^{-1}(\nabla_b \phi\delta^a_c+\nabla_c \phi\delta^a_b-\nabla^a \phi g_{bc})$ are the affine connections of a Weyl-type manifold. These do not coincide with the Christoffel symbols of the Jordan frame metric $\bf g$ and, correspondingly, one can define metric and affine magnitudes and operators in a Weyl-type manifold. Weyl-type geometry is given by the parallel transport law (2.14) and by the length transport law:

\begin{equation}
dg(\xi,\xi)=\phi^{-1}dx^n \nabla_n \phi g(\xi,\xi),
\end{equation}
that is equivalent to (2.11) in respect to the conformal transformation (2.2). All of this means that the Jordan frame formulation of general relativity should be linked with a Weyl-type geometry with units of measure varying length over the manifold according to the length transport law (2.15). In the Jordan frame GR, in particular the gravitational constant $G$ varies in spacetime like $\phi^{-1}$ while the rest masses of material particles $m$ vary like $\phi^\frac{1}{2}$, i.e. $Gm^2=const.$ is preserved. It is a conformal invariant feature of general relativity. In fact $Gm^2$ is a dimensionless constant and hence it is unchanged under the transformation of the physical units (2.2).

According to the viewpoit on the physical equivalence of the conformal formulations of a given theory of gravitation presented in \cite{iq}, both Jordan frame and Einstein frame formulations of general relativity are physically equivalent. Otherwise, these formulations of GR are equally consistent with the observational evidence. These are not alternative theories but alternative formulations of the same theory. This leads that we have two different geometrical representations of a same physical situation. In the one representation (Riemann geometry) the units of measure are constant over the manifold. In the other representation (Weyl-type geometry) the units of measure are variable over the spacetime. None of them is preferred over the other. It is a matter of mathematical convenience or, may be, philosophical prejudice which representation of the theory one chooses for the description of the given physics. This point will be further discussed in section VI.

\section{Jordan frame general relativity}

This formulation of general relativity is based on the Lagrangian (2.13). It is not a complete geometrical theory. Gravitational effects are described here by a scalar field in a Weyl-type manifold, i.e. the gravitational field shows both tensor (spin-2) and scalar (spin-0) modes. For example, in this formulation of GR the gravitational redshift effect appears only partially as a metric phenomenon, the remainder of the effect being due to a real change in the energy levels of the atoms ($m\sim\phi^\frac{1}{2}$).  

The field equations of the Jordan frame GR theory can be derived either directly from (2.13) by taking the variational derivatives of the Lagrangian respect to the dynamical variables or by conformally mapping equations (2.6) and (2.7) back to the JF metric according to (2.2). We obtain:

\begin{equation}
G_{ab}=\frac{8\pi}{\phi} T_{ab}+\frac{\omega}{\phi^2}(\nabla_a  \phi \nabla_b \phi- \frac{1}{2} g_{ab} g^{nm} \nabla_n \phi \nabla_m \phi)+\frac{1}{\phi}(\nabla_a \nabla_b \phi-g_{ab} \Box \phi),
\end{equation}
and

\begin{equation}
\Box \phi=0,
\end{equation}
where $T_{ab}=\frac{2}{\sqrt{-g}}\frac{\partial}{\partial g^{ab}}(\sqrt{-g} \phi^2 L_{matter})$ is the stress-energy tensor for ordinary matter in the Jordan frame. The energy is not conserved since the scalar field $\phi$ exchanges energy with the metric and with the matter fields. The corresponding dynamic equation is:

\begin{equation}
\nabla_n T^{na}=\frac{1}{2} \phi^{-1} \nabla^a \phi T,
\end{equation}

The equation of motion of an uncharged, spinless mass point that is acted on by the JF metric field $\bf g$ and by the scalar field $\phi$,

\begin{equation}
\frac{d^2x^a}{ds^2}=-\Gamma^a_{mn} \frac{dx^m}{ds} \frac{dx^n}{ds}-\frac{1}{2} \phi^{-1} \nabla_n \phi(\frac{dx^n}{ds}\frac{dx^a}{ds}-g^{an}),
\end{equation}
does not coincide with the geodesic equation of the Jordan frame metric. 

Most authors consider that one of the most undesirable features of the Jordan frame formulation of BD-type theories is linked with the fact that, in this frame the stress-energy tensor for the scalar field $\phi$ ($\frac{\phi}{8\pi}$ times the sum of the 2nd and 3rd terms in the right hand side(r.h.s.) of eq.(3.1)) has a non-canonical form. This leads that the scalar field kinetic energy is negative definite (or indefinite) implying that the theory may not have a stable ground state\cite{fgn}.

However, as noted in reference \cite{ss} the terms with the second covariant derivatives of the scalar field contain the connection, and hence a part of the dynamical description of gravity. In \cite{ss} a new connection was presented that leads to a canonical form of the scalar field stress-energy tensor in the Jordan frame. We obtain the same result in an alternative way. Equation (3.1) can be written in terms of affine magnitudes in the Weyl-type manifold. In this case the affine connections of the JF (Weyl-type) manifold $\gamma^a_{bc}$ do not coincide with the Christoffel symbols of the JF metric $\Gamma^a_{bc}$ (see section II). Then we can define the 'affine' Einstein tensor $^\gamma G_{ab}$ that is given in terms of the affine connections of the manifold $\gamma^a_{bc}$ instead of the Christoffel symbols of the Jordan frame metric $\Gamma^a_{bc}$. Equation (3.1) can then be rewritten as:

\begin{equation}
^\gamma G_{ab}=\frac{8\pi}{\phi} T_{ab}+\frac{(\omega+\frac{3}{2})}{\phi^2}(\nabla_a  \phi \nabla_b \phi- \frac{1}{2} g_{ab} g^{nm} \nabla_n \phi \nabla_m \phi),
\end{equation} 
where now $\frac{\phi}{8\pi}$ times the second term in the r.h.s. of this equation shows the canonical form for the scalar field stress-energy tensor. This way the main physical objection against this formulation of general relativity is removed. We shall call this as the 'true' stress-energy tensor for $\phi$ while $\frac{\phi}{8\pi}$ times the sum of the 2nd and 3rd terms in the r.h.s. of eq.(3.1) we shall call as the 'effective' stress-energy tensor for the BD scalar field $\phi$. The r.h.s. of eq.(3.1) may be negative definite implying that some energy conditions may not be fulfilled and hence the relevant singularity theorems may not hold. However in the ligth of the comments above this does not imply that the energy conditions for the 'true' matter content of the theory (the usual energy conditions applied to the sum of the stress-energy tensor for ordinary matter and the 'true' stress-energy tensor for the scalar field) do not hold. 

Another remarkable feature of the Jordan frame GR theory is that it is invariant in form under the following conformal transformations (these are in fact transformations of the units of length, time, and mass):

\begin{eqnarray}
\tilde g_{ab}&=&\phi^2 g_{ab},\nonumber\\
\tilde \phi&=&\phi^{-1},
\end{eqnarray}
and

\begin{eqnarray}
{\tilde g_{ab}}&=&f g_{ab},\nonumber\\
\tilde \phi&=&f^{-1}\phi,
\end{eqnarray}
where $f$ is some smooth function given on the manifold. The invariance can be verified by direct substitution of (3.6) or (3.7) in (2.13) or (3.1-3.4). The Lagrangian (2.13) is also invariant in respect to the more general conformal transformation\cite{iq} (first presented in \cite{far}):

\begin{eqnarray}
\tilde \gab&=&\phi^{2\alpha}\gab,\nonumber\\
\tilde \phi&=&\phi^{1-2\alpha}.
\end{eqnarray}

This transformation is accompanied by a redefinition of the BD coupling constant:

\begin{equation}
\tilde \omega=\frac{\omega-6\alpha(\alpha-1)}{(1-2\alpha)^2},
\end{equation}
with $\alpha \neq \frac{1}{2}$. The case $\alpha = \frac{1}{2}$ constitute a singularity in the transformations (3.10-3.12).

We shall point out that, for instance, Brans-Dicke theory (Lagrangian (2.4)) is invariant under (3.8,3.9) only in the absence of ordinary matter or for matter with a trace-free stress-energy tensor. The Einstein frame formulation of BD-type theories is not invariant under the conformal transformations (3.8,3.9) too.

\section{Jordan frame general relativity and the cosmological singularity}

In this section we shall study flat FRW universes given, in the Einstein frame by the line element (we use coordinates $t,r,\theta,\varphi$):

\begin{equation}
d\hat s^2=-dt^2+\hat a^2(t) (dr^2 + r^2 d\Omega^2)
\end{equation}
where $\hat a(t)$ is the Einstein frame scale factor, and $d\Omega^2\equiv d\theta^2 + sin^2 \theta d\varphi^2$. The universe is supposed to be filled with a barotropic perfect fluid ($\hat p=(\gamma-1) \hat \mu$, $\hat \mu$ is the energy density of matter and the barotropic index $0 < \gamma < 2$). We shall consider arbitrary $\omega > -\frac{3}{2}$. For $\omega<-\frac{3}{2}$ the kinetic term of the BD scalar field $\hat \phi$ in the Einstein frame has a negative energy. The case with $\omega=-\frac{3}{2}$ has been already studied in ref.\cite{iq}. The Jordan frame field equation (2.6) can be reduced to the following equation for determining the Einstein frame scale factor:

\begin{equation}
(\frac{\dot {\hat a}}{\hat a})^2=\frac{8\pi}{3} \frac{(C_2)^2}{\hat a^{3\gamma}}+\frac{1}{6}(\omega+\frac{3}{2})\frac{(C_1)^2}{\hat a^6},
\end{equation}
where $C_1$ and $C_2$ are arbitrary integration constants. While deriving (4.2) we have considered that $\hat \mu=\frac{(C_2)^2}{\hat a^{3\gamma}}$ and that after integrating eq.(2.7) once

\begin{equation}
\dot {\hat \phi}=\frac{C_1}{\hat a^3}.
\end{equation}

The overdot means derivative with respect to the Einstein frame proper time $t$. If we introduce the time variable:

\begin{equation}
dt=\frac{\hat a^{3\gamma-3}}{6-3\gamma} d\eta,
\end{equation}
then eq.(4.2) can be readily integrated to give:

\begin{equation}
\hat a(\eta)=A^{\frac{\alpha}{3}}(\eta^2-\eta_0^2)^{\frac{\alpha}{3}},
\end{equation}
where we have defined $A \equiv \frac{2\pi}{3} (C_2)^2$, $\alpha \equiv \frac{1}{2-\gamma}$, $\eta_0=\frac{3}{8\pi} \frac{C_1}{\beta (C_2)^2}$, and $\beta \equiv \frac{1}{\sqrt{\frac{2}{3}\omega+1}}$. The scalar field $\hat \phi$ can be found from eq.(4.3):

\begin{equation}
\hat \phi^{\pm}(\eta)=\hat \phi_0 \pm ln[\frac{\eta-\eta_0}{\eta+\eta_0}]^{\frac{2}{3}\alpha \beta}.
\end{equation} 

The Jordan frame scale factor $a^{\pm}(\eta)=\hat a(\eta) exp[-\frac{1}{2}\hat \phi^{\pm}(\eta)]$ is given by the following expression:

\begin{equation}
a^{\pm}(\eta)=\frac{A^{\frac{\alpha}{3}}}{\sqrt{\phi_0}}[(\eta-\eta_0)^{1\mp\beta} (\eta+\eta_0)^{1\pm\beta}]^{\frac{\alpha}{3}}. 
\end{equation}

This solution shows two branches. These are given by the choice of the '+' and the '-' signs in (4.7). It is valid for any finite $\beta > 0$ ($-\frac{3}{2} < \omega < \infty$). The relation between the proper time $\tau$ in the Jordan frame and the time variable $\eta$ is given by:

\begin{equation}
(\tau-\tau_0)^{\pm}=\frac{\alpha A^{(\gamma-1)\alpha}}{3\sqrt{\phi_0}}\int[(\eta-\eta_0)^{\gamma-1\mp\frac{\beta}{3}}(\eta+\eta_0)^{\gamma-1\pm\frac{\beta}{3}}]^\alpha d\eta.
\end{equation}

For big $\eta\gg \eta_0$,

\begin{equation}
(\tau-\tau_0)^{\pm}\approx\frac{A^{(\gamma-1)\alpha}}{3\sqrt{\phi_0}\gamma}\eta^{\alpha\gamma},
\end{equation}
so that, $\eta\rightarrow +\infty$ leads that $\tau\rightarrow +\infty$ for both '+' and '-' branches of our solution. For $\eta\rightarrow \eta_0$ we have:

\begin{equation}
(\tau-\tau_0)^{\pm}\approx\frac{A^{(\gamma-1)\alpha}}{(3\mp\beta)\sqrt{\phi_0}}[(2\eta_0)^{\gamma-1\pm\frac{\beta}{3}}(\eta-\eta_0)^{1\mp\frac{\beta}{3}}]^\alpha.
\end{equation}

In the '+' branch of the solution it is valid for any $\beta\neq 3$ ($\omega\neq -\frac{4}{3}$). For $\beta=3$ we obtain in this limit (in the '+' branch of the solution):

\begin{equation}
\tau-\tau_0\sim\frac{\alpha A^{(\gamma-1)\alpha}}{3\sqrt{\phi_0}}(2\eta_0)^{\gamma\alpha}\ln(\eta-\eta_0).
\end{equation}

Summing up. For the '-' branch we get that, when $\eta\rightarrow \eta_0$ in the Einstein frame, in its conformal Jordan frame $\tau\rightarrow \tau_0$. Then in this branch of the solution the evolution of the flat, perfect-fluid-filled, FRW universe (in the Jordan frame) is basically the same as in the Einstein frame. It evolves from a global (cosmological) singularity at the beginning of time $\tau_0$ ($a^-(\eta_0)=0$) into an infinite size universe at the infinite future $\tau=+\infty$ ($a^-(+\infty)=\infty$). Otherwise, when we work with the '+' branch of the solution, in the range $3\leq \beta<\infty$ ($-\frac{3}{2}<\omega\leq -\frac{4}{3}$), $\eta\rightarrow\eta_0$ means $\tau\rightarrow -\infty$. In this case the Jordan frame flat FRW universe evolves from an infinite size at the infinite past ($a^+(-\infty)=\infty$) into an infinite size at the infinite future ($a^+(+\infty)=\infty$), through a minimum size $a^*=\frac{1}{\sqrt{\phi_0}}A^\frac{\beta}{3} \frac{(\beta+1)^{\beta+1}}{(\beta-1)^{\beta-1}} \eta_0^{\frac{2}{3} \beta}$ at some intermediate time $\eta^*=\beta \eta_0$. In this range there is no curvature singularity, neither in the past, nor in the future.

In reference \cite{iq} the same behaviour was found in the case $\omega=-\frac{3}{2}$. This way, when we study general relativity with an extra scalar field (Lagrangians (2.5) for the Einstein frame formulation and (2.13) for the Jordan frame one) we find that there is a branch of the flat, perfect-fluid, FRW solution to the field equations of the theory where, in the range $-\frac{3}{2}\leq\omega\leq -\frac{4}{3}$, $0 < \gamma < 2$ of the free parameters $\omega$ and $\gamma$, the Jordan frame universe is free of the cosmological singularity. Unlike this in the Einstein frame representation the cosmological singularity is always present.     

We should compare this result with that obtained in ref.\cite{ps} for BD theory given by the JF Lagrangian (2.4). In this case the cosmological singularity is avoided only in some regions (regions IV and VII in fig.5 of ref.\cite{ps}) in the section $-\frac{3}{2}\leq\omega\leq -\frac{4}{3}$, $0 < \gamma < 2$ of the parameter space.

Our result serves as an illustration of the notion of geometrical duality developed in \cite{iq} (see section III), and should be interpreted (in the light of the postulate of the physical equivalence of conformal representations of a given classical theory of gravity) in the following way. Both Einstein frame picture with the cosmological singularity and the Jordan frame one without them (in the given region of the parameter space) are physically equivalent and equally consistent with the observational evidence. In the EF picture (Lagrangian (2.5)) linked with Riemann geometry, test particle's rest masses as well as the gravitational constant $G$ are real constants over the spacetime. Meanwhile, in its dual Jordan frame picture due to the Lagrangian (2.13) and linked with Weyl-type geometry, these magnitudes are not constant anymore. It is directly related to the fact that, in Weyl-type geometry, units of measure vary over the manifold. 

Although we can work (in principle) in any one of the conformal frames (JF, EF or other conformal frames), when we approach the cosmological singularity occuring in the Einstein frame formulation of general relativity, the only way we are able to describe the physics there (without renouncing to the known physical laws) is by 'jumping' to the Jordan frame representation where the singularity  is removed (in the '+' branch of the solution for the given region in the parameter space). 

The following paradox is remarkable and should be discussed. If we take the Einstein frame GR for the description of the physics, the following situation takes place. The past-directed geodesics of all of the fluid particles converge in a point of infinite density where the known physical laws break down and the very existence of these particles abruptly ends up. When we choose the Jordan frame formulation of general relativity for the description of the given physical situation we find a very unlike picture. The past-directed world-lines of the free-falling fluid particles (these do not coincide with the geodesics of the JF metric) converge up to a time in the past where the cross section area of the world-tube is a minimum and then they diverge for ever. Even if the physical interpretation of the physical rality may not be unique, the physical reality itself should be unique (it is the basic postulate in physics). The paradoxical fact is that the fluid particles are part of this reality (they are unique) and it is difficult to imagine that in one picture they have a finite life-time (into the past) while in the other they are eternal. This question is even more difficult since experiment can not help us in the search for an answer. This point will be further discussed in section VI.

\section{The low-energy limit of string theory}

Finally we shall outline some implications of the present viewpoint for the low-energy limit of string theory. It is rooted the belief that in the Planck energy scales gravity is not driven by Einstein's general relativity, but by some of its scalar-tensor modifications. In particular the low-energy theory of the fundamental string contains the BD-type theory given by the basic Lagrangian (2.1) with $\omega=-1$. Actually, for pure dilaton gravity we have:

\begin{equation}
L=\frac{\sqrt{-g}}{16\pi}e^{-2\varphi}(R + 4(\nabla\varphi)^2),
\end{equation}
where $R$ is the Ricci scalar of the four-dimensional external spacetime and $\varphi$ is the dilaton field. Under the field redefinition $\phi=e^{-2\varphi}$, the Lagrangian (5.1) can be transformed into (2.1) with $\omega=-1$. We should remember, however, that the theory given by (2.1) can be linked either with Riemann geometry or with Weyl-type geometry indistinctly (see section II) so, in this case, BD theory and general relativity with an extra scalar field are undistinguishable. This degeneration vanishes when matter fields are present. In this case dilaton gravity is given by \cite{sen}:

\begin{equation}
L=\frac{\sqrt{-g}}{16\pi}e^{-2\varphi}(R + 4(\nabla\varphi)^2)+e^{2(a-1)\varphi}L_{matter}.
\end{equation}

This Lagrangian can not be reduced to the corresponding Lagrangian (2.4) for Brans-Dicke theory with $\omega=-1$ by the redefinition $\phi=e^{-2\varphi}$, because the non-minimal coupling between the matter Lagrangian $L_{matter}$ and the dilaton field $\varphi$ in (5.2). Only for $a=1$ there is not coupling between $L_{matter}$ and the dilaton, and the Lagrangian (5.2) can be succesfully transformed into the corresponding BD Lagrangian (2.4) \cite{sen}. However, other values for $a$ ($a\neq 1$) are also available and should be taken into account. In particular, when $a=-1$, eq.(5.2) can be transformed into (2.13) with $\omega=-1$, that is the JF Lagrangian for GR theory with an extra scalar field, given in the EF by:

\begin{equation}
L=\frac{\sqrt{-\hat g}}{16\pi}(\hat R -\frac{1}{2}(\hat \nabla \hat \phi)^2)+ L_{matter}.
\end{equation}

When solitonic degrees of freedom such as p-branes are taken into account, then the effective Lagrangian can be written as BD Lagrangian (2.4) with $\omega$ given by \cite{duf}:

\begin{equation}
\omega=-\frac{(D-1)(d-2)-d^2}{(D-2)(d-2)-d^2},
 \end{equation}
where $d=p+1$. In four dimensions ($D=4$), $\omega=-\frac{4}{3}$ for the 0-brane and $\omega=-\frac{3}{2}$ for the instanton ($p=-1$).

If one assumes that, in the regime of Planck length curvature, gravity is described by general relativity with an extra scalar field, given in the Jordan frame by the Lagrangian (2.13) with $\omega$ given by (5.4) ($D=4$), and considering the gas of solitonic p-brane as a perfect fluid with the barotropic equation of state \cite{ps}, then one can conclude that the cosmological singularity can be avoided in some cases (in particular for 0-brane and for the instanton), while for the fundamental string (1-brane) the cosmological singularity is unavoidable because the value $\omega=-1$ falls outside the range $-\frac{3}{2}\leq\omega\leq -\frac{4}{3}$ (see section IV). This result should be interpreted in the light of recent developments of string theory suggesting that, in the high curvature regime, the solitonic p-brane will be copiously produced since they become light and dominate the universe in that regime \cite{ps}. Then our result is an indication that, in such a extreme regime, the cosmological singularity may be removed by the solitonic degrees of freedom such like the 0-brane and the instanton.

\section{Is the cosmological singularity really avoidable?}

The result that in the Jordan frame formulation of general relativity the cosmological singularity vanishes (in the 'plus' branch of the solution) was expected since $R_{mn}k^m k^n$ is negative definite in this frame for any non-spacelike vector $\bf k$. This means that the relevant singularity theorems may not hold. This is in contradiction with the Einstein frame formulation where $\hat R_{mn}k^m k^n$ is non-negative and a space-like singularity at the beginning of time $t=0$ ($\eta=\eta_0$) occurs. The very striking fact is that both geometrical representations with and without the cosmological singularity are observationally indistinguishable since, on the one hand they are equivalent in respect to the conformal transformation of the metric (2.2) and, on the other the physical experiment is insensible to this particular transformation of the physical units. A more careful analysis of the physical equivalence between a picture with singularity and a picture without them shows that this is a very paradoxical situation (see section IV). In fact, we can, in principle, link a physical observer with each one of the fluid particles. Suppose that a part of these observers (A observers) take the Einstein frame formulation of general relativity for the description of the evolution of the universe while the other part of observers (B observers) take the Jordan frame GR for modeling the universe. The free-fall world-lines of the A observers (their geodesic lines) meet together a finite proper time into the past. At this point the known physical laws break down and the very existence of the A observers abruptly ends up, i.e. their life-time is finite into the past. The B observers, on the contrary, are eternal (both into the past and into the future) and the physical laws they know (these are the same for both sets of observers) hold for all times. This is a very profound paradox and we do not pretend to solve it. We can only pretend to conjecture on this subject.

Two explanations to this paradoxical situation are possible. The first one is based on the fact that Einstein's theory is a classical theory of spacetime and near of the cosmological singularity we need a quantum theory (this theory has not been well stablished at present). When a viable quantum gravity theory will be worked out it may be that this singularity is removed. Our result, when applied to the low-energy limit of string theory (section V), is an indication in favour of this idea since string theory seems to be the better candidate for a final quantum theory of gravity. In the Jordan frame no singularity occurs (in the '+' branch of the solution for the given region $-\frac{3}{2}\leq\omega\leq -\frac{4}{3}$, $0 < \gamma < 2$ in the parameter space) and, consequently, we do not need of any quantum theory for describing gravitation\footnotemark\footnotetext{It is true provided the minimum value $a^*$ of the Jordan frame scale factor is much greater than the Planck length.}. This explanation is in agreement with a point of view developed in reference \cite{shojai}. According to this viewpoint, to bring in the quantum effects into the classical gravity theory one needs to make only a conformal transformation. If we start with Einstein's classical theory of gravitation then we can set in the quantum effects of matter by simply making a conformal transformation into, say, the Jordan frame. In this sense the Jordan frame formulation of general relativity already contains the quantum effects (Jordan frame GR represents a unified description of both gravity and the quantum effects of matter). 

The second possibillity is more radical. The Einstein frame formulation is not invariant under the particular transformations of the units of time, length and mass studied in section III. It is very striking since the physical laws should be invariant under the transformations of units. Unlike this The Jordan frame formulation of general relativity is invariant in respect to these transformations. This means that the picture without the cosmological singularity is more viable than the one with them. Consequently the cosmological singularity is a fictitious entity due to a wrong choice of the formulation of the theory. 

In the last instance it may be that both possibillities are connected. We hope that when a viable Einstein frame quantum theory of gravity will be worked out it should be invariant under the transformations of the physical units. All of this is, of course, a matter of pure conjecture and we hope other possibillities will be worked out. The fatal point is that this will remain a subject of conjecture since experiment can not help us in solving the present paradox.

\begin{center}
{\bf AKNOWLEDGEMENT}
\end{center}

We thank the unknown referee for recommendations and MES of Cuba by financial support.


\begin{thebibliography}{99}

\bibitem{iq} I. Quiros, gr-qc/9905071 (accepted for publication in Phys. Rev. D).
\bibitem{sok} L. M. Sokolowski in 'Proceedings of the 14th International Conference on General Relativity and Gravitation', Firenze, Italy 1995, M. Francaviglia, G. Longhi, L. Lusanna, E. Sorace (eds.) 337(World Scientific, 1997); gr-qc/9511073; G. Magnano and L. M. Sokolowski, Phys. Rev. D \textbf{50}, 5039(1994).
\bibitem{fgn} V. Faraoni, E. Gunzig and P. Nardone, IUCAA 24/98; gr-qc/9811047 (to appear in 'Fundamentals of Cosmic 
Physics').
\bibitem{ss} D. I. Santiago, A. S. Silbergleit, gr-qc/9904003 ( Submitted to Phys. Rev. D 15).
\bibitem{dk} R. H. Dicke, phys. Rev. \textbf{125}, 2163(1962).
\bibitem{am} A. Albrecht and J. Magueijo, Phys. Rev. D \textbf{59}, 043516(1999); astro-ph/9811018.
\bibitem{bdk} C. Brans and R. H. Dicke, Phys. Rev. \textbf{124}, 925(1961).
\bibitem{far} V. Faraoni, Phys. Lett. A \textbf{245}, 26(1998).
\bibitem{ps} C. Park and S-J. Sin, phys. Rev. D\textbf{57}, 4620(1998).
\bibitem{sen} A. A. Sen, gr-qc/9808075.
\bibitem{duf} M. J. Duff, R. R. Khuri and J. X. Lu, Phys. Rep. \textbf{259}, 213(1995).
\bibitem{shojai} F. Shojai, A. Shojai and M. Golshani, Mod. Phys. Lett. A. \textbf{13}, No. 34, 2725(1998); Mod. Phys. Lett. A. \textbf{13}, No. 36, 2915(1998).

\end{thebibliography}
\end{document}